\documentstyle[12pt]{article}
\newcommand{\beq}{\begin{equation}}
\newcommand{\eeq}{\end{equation}}
\newcommand{\beqa}{\begin{eqnarray}}
\newcommand{\eeqa}{\end{eqnarray}}
\newcommand{\ba}{\begin{array}}
\newcommand{\ea}{\end{array}}

\begin{document}

\begin{flushright}
Preprint CAMTP/96-5\\
July 1996\\
\end{flushright}

\vskip 0.5 truecm

\begin{center}
\large
{\bf WKB to all orders and the accuracy of the semiclassical quantization}\\
\vspace{0.25in}
\normalsize
Marko Robnik$^{(*)}$\footnote{e--mail: robnik@uni-mb.si} and 
Luca Salasnich$^{(*)(+)}$\footnote{e--mail: salasnich@math.unipd.it} \\
\vspace{0.2in}
$^{(*)}$ Center for Applied Mathematics and Theoretical Physics,\\
University of Maribor, Krekova 2, SLO--2000 Maribor, Slovenia\\
\vspace{0.2in}
$^{(+)}$ Dipartimento di Matematica Pura ed Applicata \\
Universit\`a di Padova, Via Belzoni 7, I--35131 Padova, Italy \\
and\\
Istituto Nazionale di Fisica Nucleare, Sezione di Padova,\\
Via Marzolo 8, I--35131 Padova, Italy
\end{center}

\vspace{0.3in}

\normalsize
{\bf Abstract.} We perform a systematic WKB expansion to all orders 
for a one--dimensional system with potential $V(x)=U_0/\cos^2{(\alpha x)}$. 
We are able to sum the series to the exact energy spectrum. 
Then we show that at any finite order the error of the WKB approximation 
measured in {\em the natural units of the mean energy level spacing}
does not go to zero when the quantum number goes to infinity.
Therefore we make the general conclusion that the semiclassical
approximations fail to predict the individual energy levels within a 
vanishing fraction of the mean energy level spacing. 

\vspace{0.6in}

Submitted to {\bf Journal of Physics A: Mathematical and General}
\normalsize
\vspace{0.1in}
  
\newpage

\par
In the last years many studies 
have been devoted to the transition from classical mechanics 
to quantum mechanics. These studies are motivated by the so--called 
quantum chaos (see Ozorio de Almeida 1990, 
Gutzwiller 1990, Casati and Chirikov 1995). 
An important aspect is the semiclassical quantization formula of the 
energy levels for integrable and quasi--integrable systems, 
i.e. the torus quantization initiated by Einstein (1917) and 
completed by Maslov (1972, 1981). 
As is well known, the torus quantization is just the first term 
of a certain $\hbar$-expansion, the so--called WKB expansion, 
whose higher terms 
can be calculated with a recursion formula at least for one degree systems 
(Dunham 1932, Bender, Olaussen and Wang 1977, Voros 1983). 
\par
Recently it has been observed by Prosen and Robnik (1993) and also 
Graffi, Manfredi and Salasnich (1994) 
that the leading--order semiclassical approximation fails to predict 
the individual energy levels within a vanishing fraction of the mean 
energy level spacing. This result has been shown to 
be true also for the leading (torus) semiclassical approximation  
by Salasnich and Robnik (1996).  
\par
In this paper we analyze a simple one--dimensional system 
for which we are able to perform a systematic 
WKB expansion to all orders resulting in a convergent series whose sum is 
identical to the exact spectrum. For this system we show that 
any finite order WKB (semiclassical) approximation 
fails to predict the individual energy levels within a vanishing
fraction of the mean energy level spacing. 
\par
The Hamiltonian of the system is given by
\beq
H={p^2\over 2m}+ V(x) \; ,
\eeq
where
\beq
V(x)= {U_0\over \cos^2{(\alpha x)} } \; .
\eeq
Of course, the Hamiltonian is a constant 
of motion, whose value is equal to the total energy $E$. 
To perform the torus quantization it is necessary to introduce 
the action variable
\beq
I={1\over 2\pi}\oint p dx = {\sqrt{2m}\over \alpha} (\sqrt{E} - \sqrt{U_0}) 
\; .
\eeq
The Hamiltonian as a function of the action reads
\beq
H={\alpha^2 \over 2m}I^2 + {2\alpha \sqrt{U_0\over 2m}} I + U_0 \; ,
\eeq
and after the torus quantization
\beq
I = (\nu + {1\over 2})\hbar \; ,
\eeq
where $\nu = 0,1,2,\dots$, the energy spectrum is given by
\beq
E_{\nu}^{tor} = A[(\nu +{1\over 2})+{1\over 2}B]^2 \; ,
\eeq
where $A=\alpha^2 \hbar^2 /(2m)$ and $B=\sqrt{8mU_0}/(\alpha \hbar )$. 
\par 
The Schr\"odinger equation of the system
\beq
[-{\hbar^2 \over 2 m} {d^2 \over dx^2} + V(x)] \psi (x) = E \psi (x) \; ,
\eeq
can be solved analytically (as shown in Landau and Lifshitz 1973, 
Fl\"ugge 1971) and the exact energy spectrum is:
\beq
E_{\nu}^{ex} = A [( \nu + {1\over 2}) +{1\over 2}\sqrt{1+B^2}]^2 \; ,
\eeq 
where $\nu = 0,1,2,\dots$. 
We see that the torus quantization does not give the correct 
energy spectrum, but it is well known that the torus quantization 
is just the first term of the WKB expansion. 
To calculate all the terms of the WKB expansion we observe that 
the wave function can always be written as
\beq
\psi (x) = \exp{ \big( {i\over \hbar} \sigma (x) \big) } \; ,
\eeq
where the phase $\sigma (x)$ is a complex function that satisfies 
the differential equation
\beq
\sigma{'}^2(x) + ({\hbar \over i}) \sigma{''}(x) = 2m(E - V(x)) \; .
\eeq
The WKB expansion for the phase is given by
\beq
\sigma (x) = \sum_{k=0}^{\infty} ({\hbar \over i})^k \sigma_k(x) \; .
\eeq
Substituting (11) into (10) and comparing like powers of $\hbar$ gives 
the recursion relation ($n>0$)
\beq
\sigma{'}_0^2=2m(E-V(x)), \;\;\;\; 
\sum_{k=0}^{n} \sigma{'}_k\sigma{'}_{n-k}
+ \sigma{''}_{n-1}= 0 \; .
\eeq
\par
The quantization condition is obtained by requiring 
the single-valuedness of the wave function
\beq
\oint d\sigma = 
\sum_{k=0}^{\infty} ({\hbar \over i})^{k} \oint d\sigma_{k}=
2 \pi \hbar \; \nu  \;\; ,
\eeq
where $\nu =0,1,2,\dots$ is the quantum number. 
\par
The zero order term, which gives the Bohr-Sommerfeld formula, 
is given by
\beq
\oint d\sigma_0 = 2 \int dx \sqrt{2m(E - V(x))} = 2\pi \hbar 
(\sqrt{E\over A}-{1\over 2}B) \; ,
\eeq
and the first odd term in the series gives the Maslov corrections
(Maslov index is equal to 2) 
\beq
({\hbar \over i}) \oint d\sigma_{1} = ({\hbar \over i}) {1\over 4} 
\ln{p}|_{contour} = - \pi \hbar \; . 
\eeq
The zero and first order terms give the equation (6), which is 
the torus quantization formula for the energy levels 
(Bohr--Sommerfeld--Maslov). 
Here we want to analyze the quantum corrections to this formula. 
We observe that all the other odd terms vanish when integrated along the closed 
contour because they are exact differentials (Bender, Olaussen and
Wang 1977). So the quantization condition (13) can be written
\beq
\sum_{k=0}^{\infty} ({\hbar \over i})^{2k} \oint d\sigma_{2k} = 2 \pi \hbar 
(\nu +{1\over 2}) \; ,
\eeq 
thus again a sum over even--numbered terms only. 
The next two non--zero terms are (Narimanov 1995, Bender, Olaussen and 
Wang 1977, Robnik and Salasnich 1996)
\beq
({\hbar \over i})^{2} \oint d\sigma_2 
= - {\hbar^2 \over \sqrt{2m}}{1\over 12} {\partial^2 \over \partial E^2} 
\int dx {V{'}^2(x) \over \sqrt{E - V(x)} } \; , 
\eeq
\beq
({\hbar \over i})^{4} \oint d\sigma_4 = {\hbar^4\over (2m)^{3/2}}
[{1\over 120} {\partial^3\over \partial E^3} 
\int dx {V{''}^2(x) \over \sqrt{E - V(x)} } 
- {1\over 288 } {\partial^4\over \partial E^4} 
\int dx {V{'}^2(x) V{''}(x) \over \sqrt{E - V(x)} } ] \; .
\eeq
A straightforward calculation of these terms gives (see the Appendix)
\beq
({\hbar \over i})^{2} \oint d\sigma_2 = - {2\pi \hbar \over 4 B} \; ,
\eeq
and
\beq
({\hbar \over i})^{4} \oint d\sigma_4 = {2\pi \hbar \over 16 B^3} \; .
\eeq
Up to the fourth order in $\hbar \sim B^{-1}$ the quantization condition reads
\beq
E_{\nu}^{(4)} = A[(\nu 
+{1\over 2})+{1\over 2}B + {1\over 4 B} - {1\over 16 B^3}]^2 
\; .
\eeq
The first two terms on the right side give the torus quantization formula, 
and the other two terms are quantum corrections. 
Higher--order quantum corrections quickly increase in complexity but in this 
specific case they can be calculated. 
We first verify by induction, following Bender, Olaussen and Wang (1977), 
that the solution to (12) has the general form
\beq
\sigma_n^{'}(x) = (\sigma_0^{'})^{1-3n}P_n(\cos{(\alpha x)}) 
\sin^{f(n)}{(\alpha x)} \; ,
\eeq
where $f(n)=0$ for $n$ even and $f(n)=1$ for $n$ odd, and $P_n$ is 
a polynomial given by
\beq
P_n(\cos{(\alpha x)}) = \sum_{l=0}^{g(n)} C_{n,l}\cos^{2l-3n}{(\alpha x)} 
\; ,
\eeq
with $g(n)=(3n-2)/2$ for $n$ even and $g(n)=(3n-3)/2$ for $n$ odd. \\
The integrals in (16) are performed by substituting $z=\tan{(\alpha x)}$. 
In this way the $2k$-term reduces to
\beq
({\hbar \over i})^{2k} \oint d\sigma_{2k} = 
({\hbar \over i})^{2k} {(2m)^{1/2-3k}\over \alpha} 
\sum_{l=0}^{3k-1} C_{2k,l} \oint dz { (1+z^2)^{3k-l-1}\over 
(E-U_0-U_0z^2)^{3k -1/2} } \; .
\eeq
We observe that
$$
\oint dz { (1+z^2)^{3k-l-1}\over (E-U_0-U_0z^2)^{3k -1/2} } = 
$$
\beq
=(-1)^{3k-1} {\Gamma ({1\over 2})\over \Gamma (3k -{1\over 2})} 
{\partial^{3k -1}\over \partial E^{3k-1} } 
\oint dz { (1+z^2)^{3k-l-1}\over (E-U_0-U_0z^2)^{1/2} } \; ,
\eeq
so the only non--zero term is for $l=0$ 
$$
{\partial^{3k -1}\over \partial E^{3k-1} } 
\oint dz { (1+z^2)^{3k-1}\over (E-U_0-U_0z^2)^{1/2} } = 
{2^{6k-1}\over U_0^{1/2}} {\Gamma (3k -1/2)^2\over \Gamma (6k -1)}
{\partial^{3k -1}\over \partial E^{3k-1} } \beta^{3k-1} \; =
$$
\beq
= {2^{6k-1}\over U_0^{1/2}}{\Gamma (3k -1/2)^2\over \Gamma (6k -1)} 
\Gamma (3k ) {1\over U_0^{3k-1/2}} 2\pi \; ,
\eeq
where $\beta = (E-U_0)/U_0$. At this stage we obtain 
\beq
({\hbar \over i})^{2k} \oint d\sigma_{2k} = (-1)^{5k-1} 
\hbar^{2k} {(2m)^{1/2-3k}\over \alpha} C_{2k,0} {1\over U_0^{3k-1/2}} 2\pi 
\; .
\eeq
\par
Now we need to find the coefficient $C_{2k,0}$ explicitly. 
By inserting (22) with (23) in the recursion relation (12) we obtain
\beq
\sum_{k=0}^n C_{k,0}C_{n-k,0} -(2mU_0\alpha ) C_{n-1,0} = 
\sum_{k=1}^{n-1} C_{k,0}C_{n-k,0} + 2C_{n,0} -(2mU_0\alpha ) C_{n-1,0} = 0 
\; ,
\eeq
from which we have
\beq
C_{k,0}={1\over 2}\big[ (2m\alpha U_0)C_{k-1,0}-\sum_{j=1}^{k-1}
C_{j,0} C_{k-j,0} \big] , \;\;\; C_{0,0}=1 \; . 
\eeq
From this equation one shows $C_{1,0}=m \alpha U_0$. Further, 
it easy to show that all higher odd coefficients vanish, 
$C_{2k+3,0}=0$ for $k=0,1,2,\dots$. 
The solution of this equation for the remaining nonzero 
even coefficients is given by
\beq
C_{2k,0}= (-1)^k (2mU_0\alpha )^{2k} 2^{-2k} {{1\over 2}\choose k} \; ,
\eeq
which can be verified by direct substitution in equation (29) 
resulting in an identity for half integer binomial coefficients. 
Then the integral (27) can be written 
\beq
({\hbar \over i})^{2k} \oint d\sigma_{2k} = (-1) \hbar^{2k} 2\pi 
\alpha^{2k-1} (2m)^{1/2-k} 2^{-2k} {{1\over 2}\choose k}U_0^{k-1/2} 
= - {1\over 2}{{1\over 2}\choose k}{2\pi \hbar \over B^{2k-1}} \; .
\eeq
In conclusion, the WKB quantization to all orders (16) is
\beq
E_{\nu}^{(\infty )} = 
A\big[ (\nu +{1\over 2})+ {1\over 2} \sum_{k=0}^{\infty} {{1\over 2}\choose k} 
{1\over B^{2k -1}} \big]^2 \; .
\eeq
Because $\sum_{k=0}^{\infty} {{1\over 2}\choose k} 
B^{1-2k} = \sqrt{1+B^2}$ we have 
$E^{ex}_{\nu}=E^{(\infty )}_{\nu}$, i.e. the WKB series converges 
to the exact result (8). 
\par
Now we can calculate the error in units of the mean level spacing 
$\Delta E_{\nu}=E^{ex}_{\nu +1}-E^{ex}_{\nu}$ 
between the exact level $E^{ex}_{\nu}$ 
and its WKB approximation $E^{(N)}_{\nu}$ to $N$th order: 
\beq
{E^{ex}_{\nu}-E^{(N)}_{\nu} \over \Delta E_{\nu}}=
{1\over 2} \sum_{k=N+1}^{\infty} 
{{1\over 2}\choose k} {1\over B^{2k-1}}
\; , \;\;\;\;\;\; \hbox{for} \;\; \nu \to \infty \; . 
\eeq
The limit clearly shows 
that even for arbitrarily small but finite $\hbar$ ($1<<B < \infty$), 
the relative error for any finite WKB approximation becomes constant
on increasing $\nu$, and scales as
\beq
{E^{ex}_{\nu}-E^{(N)}_{\nu} \over \Delta E_{\nu} } \sim {1\over 2} 
{{1\over 2} \choose N+1} {1\over B^{2N+1}} \; , \;\;\;\;\;\; B\to \infty \; .
\eeq
Note that the limit $B\to \infty$ is equivalent to the limit $\hbar \to 0$. 
\\\\
For our present system we can conclude that to 
any finite order semiclassical approximation the error measured 
in units of the mean level spacing remains constant even 
if the quantum number increases indefinitely, 
contrary to the naive expectation. This confirms the general 
statements made by Prosen and Robnik (1993). 
We have thus provided a clear demonstration 
that the semiclassical methods cannot predict the
individual energy levels (and also their wavefunctions)
within a vanishing fraction of the mean energy level spacing.
Therefore we cannot expect the semiclassics to correctly describe
the fine structure of energy spectra manifested in the 
short range statistics like the energy level repulsion,
which was predicted to be a purely quantum effect (Robnik 1986),
later reconfirmed by Berry (1991). On the other hand Prosen
and Robnik (1993) have shown that the long range statistics
of the energy spectra are very well captured even by the lowest
order semiclassical approximation. This is of course compatible
with the very important semiclassical theory of delta 
statistics $\Delta(L)$ (spectral rigidity) by Berry (1985), employing the
Gutzwiller periodic orbit theory (1990), where agreement with
predictions of random matrix theories and with the experimental
and numerical data has been obtained at large $L$. Also,
Berry and Tabor (1977) have used torus quantization of integrable
systems (with many degrres of freedom), predicting the Poissonian
(exponential) energy level distribution. Our results show that their
result cannot be rigorous, especially as we know some counterexamples
of integrable systems with non-Poissonian statistics (Bleher {\em et al}
1993), and also know that their approximation does not take into 
account the nonperturbative tunneling effects, but it is nevertheless
a heuristic argument explaining why {\em typically} we do observe
Poissonian statistics in classically integrable systems. By
typically we mean that the set of exceptions has a small or even
vanishing measure.

\par
The conclusion of this paper is that the semiclassical methods are
just not good enough (at any order) to describe the fine structure
of energy spectra and wavefunctions. 
Our approach leading to the above conclusion rests upon 
a systematic WKB expansion for the potential $V(x)=U_0/\cos^2{(\alpha x)}$ 
using the technique of Bender, Olaussen and Wang (1977). 
We are able to calculate all orders, the series is convergent 
and can be summed precisely to the exact result.  

\section*{Acknowledgements}
\par
LS acknowledges the Alps-Adria Rectors Conference Grant of 
the University of Maribor. 
MR thanks Dr. Evgueni Narimanov and Professor Douglas A. Stone
(Yale University) for stimulating discussions and for communicating
related results. The financial support by the Ministry of Science
and Technology of the Republic of Slovenia is acknowledged with
thanks.

\newpage

\section*{Appendix}

In this appendix we show how to obtain the formulas (19) and (20). 
In all integrals of this section the limits of integration are
between the two turning points. 
After substitution $z=\tan{(\alpha x)}$, we have
\beqa
\int dx {V{'}^2(x) \over \sqrt{E - V(x)} } & = & 
{4 \alpha U_0^2\over \sqrt{U_0}} 
\int_{-\sqrt{\beta}}^{\sqrt{\beta}} dz {z^2 (z^2+1)\over \sqrt{\beta - z^2}} =
\nonumber 
\\
& = & {4 \alpha U_0^2\over \sqrt{U_0}} (3 \beta^2 + 4\beta ){\pi\over 8}) 
\; ,
\eeqa
where $\beta = (E-U_0)/U_0$. In conclusion we have
\beq
({\hbar \over i})^2 \oint d\sigma_2 = - {\hbar^2 
\alpha \pi \over 8 \sqrt{2 m U_0}}=- {2\pi \hbar \over 4 B} \; ,
\eeq
with $B=\sqrt{8mU_0}/(\alpha \hbar )$. 
\par
To obtain the formula (21) we proceed in the same way.  
\beqa
\int dx {V{''}^2(x) \over \sqrt{E - V(x)} } & = & 
{4 \alpha^3 U_0^2\over \sqrt{U_0}} 
\int_{-\sqrt{\beta}}^{\sqrt{\beta}} dz 
{(9z^4+6z^2+1)(z^2+1)\over \sqrt{\beta - z^2}} =
\nonumber
\\
& = & {4 \alpha U_0^2\over \sqrt{U_0}} (45 \beta^3 +90 \beta^2 +56 \beta + 16 )
{\pi\over 16} \; , 
\eeqa
From which we obtain
\beq
{\partial^3\over \partial E^3} 
\int dx {V{''}^2(x) \over \sqrt{E - V(x)} } = 
{135 \pi \alpha^3 \sqrt{U_0}\over 2 U_0^2} \; .
\eeq
For the last integral we have
\beqa
\int dx {V{'}^2(x) V{''}(x) \over \sqrt{E - V(x)} } & = & 
{8 \alpha^3 U_0^2\over \sqrt{U_0}} 
\int_{-\sqrt{\beta}}^{\sqrt{\beta}} dz 
{z^2(3z^2+1)(z^2+1)^2\over \sqrt{\beta - z^2}} =
\nonumber
\\
& = & {8 \alpha U_0^2\over \sqrt{U_0}} (105 \beta^4 + 280 \beta^3 + 
240 \beta^2 + 60 \beta ){\pi\over 128} \; ,
\eeqa
from which we obtain
\beq
{\partial^4\over \partial E^4} 
\int dx {V{'}^2(x) V{''}(x) \over \sqrt{E - V(x)} }  
={315 \pi \alpha^3 \sqrt{U_0}\over 2 U_0^2} \; .
\eeq
In conclusion we have
\beqa
({\hbar \over i})^4 \oint d\sigma_4 & = & {\hbar^4 \over (2m)^{3/2}}
{\alpha^3 \sqrt{U_0}\over U_0^2} \big[ 
{1\over 120} {135 \over 2} - {1\over 288} {315 \over 2} \big]= 
\nonumber
\\
& = & {\hbar^4 \alpha^3 \pi 
\sqrt{U_0}\over 64 (2m)^{3/2}U_0^2} = {2\pi \hbar \over 16 B^3} \; .
\eeqa

\newpage

\section*{References} 
\parindent=0. pt

Bender C M, Olaussen K and Wang P S 1977 {\it Phys. Rev. } D {\bf 16} 1740 
\\\\
Berry M V 1985 {\it Proc. Roy. Soc. London A} {\bf 400} 229
\\\\
Berry M V 1991 in {\it Chaos and Qunatum Physics}, eds. M.-J. Giannoni,
A. Voros and J. Zinn-Justin (Amsterdam: North-Holland) 251
\\\\
Berry M V and Tabor M 1977 {\it Proc. Roy. Soc. London A} {\bf 356} 375
\\\\
Bleher P M, Cheng Zheming, Dyson F J and Lebowitz J L 1993
{\it Commun. Math. Phys.} {\bf 154} 433
\\\\
Casati G and Chirikov B V 1995 {\it Quantum Chaos} (Cambridge: Cambridge 
University Press) 
\\\\
Dunham J L 1932 {\it Phys. Rev.} {\bf 41} 713 
\\\\
Einstein A 1917 {\it Verh. Dtsch. Phys. Ges.} {\bf 19} 82
\\\\
Fl\"ugge S 1971 {\it Practical Quantum Mechanics I} (Berlin: Springer)
\\\\
Graffi S, Manfredi V R and Salasnich L 1994 
{\it Nuovo Cim.} B {\bf 109} 1147 
\\\\
Gutzwiller M C 1990 {\it Chaos in Classical and Quantum Mechanics} 
(New York: Springer) 
\\\\
Landau L D and Lifshitz E M 1973 {\it Nonrelativistic Quantum Mechanics} 
(Moscow: Nauka) 
\\\\
Maslov V P 1961 {\it J Comp. Math. and Math. Phys.} {\bf 1} 113--128; 
638--663 (in Russian)
\\\\
Maslov V P and Fedoriuk M V 1981 {\it Semi-Classical Approximations in 
Quantum Mechanics} (Boston: Reidel Publishing Company), and the references 
therein 
\\\\
Narimanov E 1995, private communication
\\\\
Ozorio de Almeida A 1990 {\it Hamiltonian Systems: Chaos and Quantization} 
(Cambridge: Cambridge University Press)
\\\\
Prosen T and Robnik M 1993 {\it J. Phys.} A {\bf 26} L37 
\\\\
Robnik M 1986 {\it Lecture Notes in Physics} {\bf 263} 120
\\\\
Robnik M and Salasnich L 1996 "WKB Expansion for the Angular Momentum 
and the Kepler Problem: from the Torus Quantization to the Exact One", 
Preprint University of Maribor, CAMTP/96-4 
\\\\
Salasnich L and Robnik M 1996 "Quantum Corrections to the Semiclassical 
Quantization of a Nonintegrable System", Preprint University of Maribor, 
CAMTP/96-1
\\\\
Voros A 1983 {\it Ann. Inst. H. Poincar\'e} A {\bf 39} 211 

\end{document}